\documentclass[titlepage,12pt]{article}
\usepackage{amssymb}
\usepackage{amsfonts}
\usepackage{amsmath}
\usepackage[dvips]{graphicx}
\usepackage{subfigure}
\usepackage[font=small,format=plain,labelfont=bf,up,textfont=it,up]{caption}
\usepackage{appendix}

\begin{document}

\title{Scattering and bound states of fermions in a mixed vector-scalar
smooth step potential}
\date{}
\author{W.M. Castilho\thanks{%
E-mail address: castilho.w@gmail.com (W.M. Castilho)} and A.S. de Castro%
\thanks{%
E-mail address: castro@pq.cnpq.br (A.S. de Castro)} \\
\\
UNESP - Campus de Guaratinguet\'{a}\\
Departamento de F\'{\i}sica e Qu\'{\i}mica\\
12516-410 Guaratinguet\'{a} SP - Brazil }
\date{}
\maketitle

\begin{abstract}
The scattering of a fermion in the background of a smooth step potential is
considered with a general mixing of vector and scalar Lorentz structures
with the scalar coupling stronger than or equal to the vector coupling.
Charge-conjugation and chiral-conjugation transformations are discussed and
it is shown that a finite set of intrinsically relativistic bound-state
solutions appears as poles of the transmission amplitude. It is also shown
that those bound solutions disappear asymptotically as one approaches the
conditions for the realization of the so-called spin and pseudospin
symmetries in a four-dimensional space-time.
\end{abstract}

\section{Introduction}

The solutions of the Dirac equation with vector and scalar potentials can be
classified according to an SU(2) symmetry group when the difference between
the potentials, or their sum, is a constant. The near realization of these
symmetries may explain degeneracies in some heavy meson spectra (spin
symmetry) \cite{pag}-\cite{gin} or in single-particle energy levels in
nuclei (pseudospin symmetry) \cite{gin}-\cite{ps}. When these symmetries are
realized, the energy spectrum does not depend on the spinorial structure,
being identical to the spectrum of a spinless particle \cite{spin0}. In
fact, there has been a continuous interest for solving the Dirac equations
in the four-dimensional space-time as well as in lower dimensions for a
variety of potentials and couplings. A few recent works have been devoted to
the investigation of the solutions of the Dirac equation by assuming that
the vector potential has the same magnitude as the scalar potential \cite%
{gum}-\cite{jia} whereas other works take a more general mixing \cite{asc1}-%
\cite{aop}.

In a recent work the scattering a fermion in the background of a sign
potential has been considered with a general mixing of vector and scalar
Lorentz structures with the scalar coupling stronger than or equal to the
vector coupling \cite{aop}. It was shown that a special unitary
transformation preserving the form of the current decouples the upper and
lower components of the Dirac spinor. Then the scattering problem was
assessed under a Sturm-Liouville perspective. Nevertheless, an isolated
solution from the Sturm-Liouville perspective is present. It was shown that,
when the magnitude of the scalar coupling exceeds the vector coupling, the
fermion under a strong potential can be trapped in a highly localized region
without manifestation of Klein's paradox. It was also shown that this
curious lonely bound-state solution disappears asymptotically as one
approaches the conditions for the realization of \textquotedblleft spin and
pseudospin symmetries\textquotedblright .

The purpose of the present paper is to generalize the previous work to a
smoothed out form of the sign potential. We consider a smooth step potential
behaving as $V(x)\sim \tanh \gamma x$. This form for the potential, termed
kink-like potential just because it approaches a nonzero constant value as $%
x\rightarrow +\infty $ and $V\left( -\infty \right) =-V\left( +\infty
\right) $, has already been considered in the literature in nonrelativistic
\cite{flu} and relativistic \cite{cont1}-\cite{cont2} contexts. The
satisfactory completion of this task has been alleviated by the use of
tabulated properties of the hypergeometric function. A peculiar feature of
this potential is the absence of bound states in a nonrelativistic approach
because it gives rise to an ubiquitous repulsive potential. Our problem is
mapped into an exactly solvable Sturm-Liouville problem of a Schr\"{o}%
dinger-like equation with an effective Rosen-Morse potential which has been
applied in discussing polyatomic molecular vibrational states \cite{ros}.
The scattering problem is assessed and the complex poles of the transmission
amplitude are identified. In that process, the problem of solving a
differential equation for the eigenenergies corresponding to bound-state
solutions is transmuted into the solutions of a second-degree algebraic
equation. It is shown that, in contrast to the case of a sign step potential
of Ref. \cite{aop}, the spectrum consists of a finite set of bound-state
solutions. An isolated solution from the Sturm-Liouville perspective is also
present. With this methodology the whole relativistic spectrum is found, if
the particle is massless or not. Nevertheless, bounded solutions do exist
only under strict conditions. Interestingly, all of those bound-state
solutions tend to disappear as the conditions for \textquotedblleft spin and
pseudospin symmetries\textquotedblright\ are approached. We also consider
the limit where the smooth step potential becomes the sign step potential.

\section{Scalar and vector potentials in the Dirac equation}

The Dirac equation for a fermion of rest mass $m$ reads%
\begin{equation}
\left( \gamma ^{\mu }p_{\mu }-Imc-V/c\right) \Psi =0  \label{d1}
\end{equation}%
where $p_{\mu }=i\hbar \partial _{\mu }$ is the momentum operator, $c$ is
the velocity of light, $I$ is the unit matrix, and the square matrices $%
\gamma ^{\mu }$ satisfy the algebra $\{\gamma ^{\mu },\gamma ^{\nu
}\}=2Ig^{\mu \nu }$. In 1+1 dimensions $\Psi $ is a 2$\times $1 matrix and
the metric tensor is $g^{\mu \nu }=$ diag$\left( 1,-1\right) $. For vector
and scalar interactions the matrix potential is written as%
\begin{equation}
V=\gamma ^{\mu }A_{\mu }+IV_{s}
\end{equation}%
We say that $A_{\mu }$ and $V_{s}$ are the vector and scalar potentials,
respectively, because the bilinear forms $\bar{\Psi}\gamma ^{\mu }\Psi $ and
$\bar{\Psi}I\Psi $ behave like vector and scalar quantities under a Lorentz
transformation, respectively. Eq. (\ref{d1}) can be written in the form
\begin{equation}
i\hbar \frac{\partial \Psi }{\partial t}=H\Psi   \label{b}
\end{equation}%
with the Hamiltonian given as%
\begin{equation}
H=\gamma ^{5}c\left( p_{1}+\frac{A_{1}}{c}\right) +IA_{0}+\gamma ^{0}\left(
mc^{2}+V_{s}\right)
\end{equation}%
where $\gamma ^{5}=\gamma ^{0}\gamma ^{1}$. Requiring $\left( \gamma ^{\mu
}\right) ^{\dag }=\gamma ^{0}\gamma ^{\mu }\gamma ^{0}$ and defining the
adjoint spinor $\bar{\Psi}=\Psi ^{\dagger }\gamma ^{0}$, one finds the
continuity equation $\partial _{\mu }J^{\mu }=0$, where the conserved
current is $J^{\mu }=c\bar{\Psi}\gamma ^{\mu }\Psi $. The positive-definite
function $J^{0}/c=|\Psi |^{2}$ is interpreted as a position probability
density and its norm is a constant of motion. This interpretation is
completely satisfactory for single-particle states \cite{tha}. If the
potentials are time independent one can write $\Psi \left( x,t\right) =\psi
\left( x\right) \exp \left( -iEt/\hbar \right) $ in such a way that the
time-independent Dirac equation becomes $H\psi =E\psi $. Meanwhile $J^{\mu }$
is time independent and $J^{1}$ is uniform. The space component of the
vector potential can be gauged away by defining a new spinor just differing
from the old by a phase factor so that we can consider $A_{1}=0$ without
loss of generality. From now on, we use the explicit representation $\gamma
^{0}=\sigma _{3}$ and $\gamma ^{1}=i\sigma _{2}$ in such a way that $\gamma
^{5}=\sigma _{1}$. Here, $\sigma _{1}$, $\sigma _{2}$ and $\sigma _{3}$
stand for the Pauli matrices. The charge-conjugation operation is
accomplished by the transformation $\psi _{c}=\sigma _{1}\psi ^{\ast }$
followed by $A_{0}\rightarrow -A_{0}$, $V_{s}\rightarrow V_{s}$ and $%
E\rightarrow -E$ \cite{prc}. As a matter of fact, $A_{0}$ distinguishes
fermions from antifermions but $V_{s}$ does not, and so the spectrum is
symmetrical about $E=0$ in the case of a pure scalar potential. The
chiral-conjugation operation $\gamma ^{5}\psi $ (according to Ref. \cite{wat}%
) is followed by the changes of the signs of $V_{s}$ and $m,$ but not of $%
A_{0}$ and $E$ \cite{prc}. One sees that the charge-conjugation and the
chiral-conjugation operations interchange the roles of the upper and lower
components of the Dirac spinor. For weak potentials, fermions (antifermions)
are subject to the effective potential $V_{s}+A_{0}$ ($V_{s}-A_{0}$) with
energy $E\approx +mc^{2}$ ($-mc^{2}$) so that a mixed potential with $%
A_{0}=-V_{s}$ ($A_{0}=+V_{s}$) is associated with free fermions
(antifermions) in a nonrelativistic regime \cite{aop}.

Introducing the unitary operator

\begin{equation}
U(\theta )=\exp \left( -\frac{\theta }{2}i\sigma _{1}\right)  \label{2}
\end{equation}

\noindent where $\theta $ is a real quantity such that $0\leq \theta \leq
\pi $, one can write $h\phi =E\phi $, where $\phi =U\psi $ and $h=UHU^{-1}$
takes the form

\begin{eqnarray}
h &=&\sigma _{1}cp_{1}+IA_{0}  \notag \\
&&  \notag \\
&&+\sigma _{3}\left( mc^{2}+V_{s}\right) \cos \theta -\sigma _{2}\left(
mc^{2}+V_{s}\right) \sin \theta  \label{333333}
\end{eqnarray}%
It is instructive to note that the transformation preserves the form of the
current in such a way that $J^{\mu }=c\bar{\Phi}\gamma ^{\mu }\Phi $. An
additional important feature of the continuous chiral transformation (see,
e.g., \cite{tou})) induced by (\ref{2}) is that it is a symmetry
transformation when $m=V_{s}=0$. In terms of the upper and the lower
components of the spinor $\phi $, \noindent the Dirac equation decomposes
into:
\begin{equation}
\hbar c\frac{d\phi _{\pm }}{dx}\pm \left( mc^{2}+V_{s}\right) \sin \theta
\,\phi _{\pm }=i\left[ E\pm \left( mc^{2}+V_{s}\right) \cos \theta -A_{0}%
\right] \phi _{\mp }  \label{eq1}
\end{equation}

\noindent Furthermore,%
\begin{equation}
\frac{J^{0}}{c}=|\phi _{+}|^{2}+|\phi _{-}|^{2},\quad \frac{J^{1}}{c}=2\text{%
Re}\left( \phi _{+}^{\ast }\phi _{-}\right)
\end{equation}

\noindent

\noindent Choosing
\begin{equation}
A_{0}=V_{s}\cos \theta  \label{5}
\end{equation}%
\noindent one has
\begin{subequations}
\begin{eqnarray}
\hbar c\frac{d\phi _{+}}{dx}+\left( mc^{2}+V_{s}\right) \sin \theta \,\phi
_{+} &=&i\left( E+mc^{2}\cos \theta \right) \phi _{-}  \label{6a} \\
&&  \notag \\
\hbar c\frac{d\phi _{-}}{dx}-\left( mc^{2}+V_{s}\right) \sin \theta \,\phi
_{-} &=&i\left[ E-\left( mc^{2}+2V_{s}\right) \cos \theta \right] \phi _{+}
\label{6b}
\end{eqnarray}%
Note that due to the constraint represented by (\ref{5}), the vector and
scalar potentials have the very same functional form and the parameter $%
\theta $ in (\ref{2}) measures the dosage of vector coupling in the
vector-scalar admixture in such a way that $|V_{s}|\geq $ $|A_{0}|$. Note
also that when the mixing angle $\theta $ goes from $\pi /2-\varepsilon $ to
$\pi /2+\varepsilon $ the sign of the spectrum undergoes an inversion under
the charge-conjugation operation whereas the spectrum of a massless fermion
is invariant under the chiral-conjugation operation. Combining
charge-conjugation and chiral-conjugation operations makes the spectrum of a
massless fermion to be symmetrical about $E=0$ in spite of the presence of
vector potential.

We now split two classes of solutions depending on whether $E$ is equal to
or different from $-mc^{2}\cos \theta $.

\subsection{$E=-mc^{2}\cos \protect\theta $}

Defining $v\left( x\right) =\int^{x}dy\,V_{s}\left( y\right) $, the
solutions for (\ref{6a}) and (\ref{6b}) with $E=-mc^{2}\cos \theta $ are
\end{subequations}
\begin{subequations}
\begin{eqnarray}
\phi _{+} &=&N_{+}  \label{ii1a} \\
&&  \notag \\
\phi _{-} &=&N_{-}-2\frac{i}{\hbar c}N_{+}\left[ mc^{2}x+v\left( x\right) %
\right] \cos \theta   \label{ii1b}
\end{eqnarray}%
for $\sin \theta =0$, and
\end{subequations}
\begin{subequations}
\begin{eqnarray}
\phi _{+} &=&N_{+}\exp \left\{ -\frac{\sin \theta }{\hbar c}\left[
mc^{2}x+v\left( x\right) \right] \right\}   \label{ii2a} \\
&&  \notag \\
\phi _{-} &=&N_{-}\exp \left\{ +\frac{\sin \theta }{\hbar c}\left[
mc^{2}x+v\left( x\right) \right] \right\} +i\phi _{+}\cot \theta
\label{ii2b}
\end{eqnarray}%
for $\sin \theta \neq 0$. \noindent $N_{+}$ and $N_{-}$ are normalization
constants. It is instructive to note that there is no solution for
scattering states. Both set of solutions present a space component for the
current equal to $J^{1}=2c\text{Re}\left( N_{+}^{\ast }N_{-}\right) $ and a
bound-state solution demands $N_{+}=0$ or $N_{-}=0$, because $\phi _{+}$ and
$\phi _{-}$ are square-integrable functions vanishing as $|x|\rightarrow
\infty $. There is no bound-state solution for $\sin \theta =0$, and for $%
\sin \theta \neq 0$ the existence of a bound state solution depends on the
asymptotic behaviour of $v(x)$ \cite{as1}, \cite{hot}. Note also that
\end{subequations}
\begin{equation}
\phi _{\pm }=N_{\pm }\exp \left\{ \mp \frac{1}{\hbar c}\left[
mc^{2}x+v\left( x\right) \right] \right\}   \label{sc1}
\end{equation}%
in the case of a pure scalar coupling ($E=0$) so that either $\phi _{+}=0$
or $\phi _{-}=0$.\noindent

\subsection{$E\neq -mc^{2}\cos \protect\theta $}

For $E\neq -mc^{2}\cos \theta $, using the expression for $\phi _{-} $
obtained from (\ref{6a}), viz.

\begin{equation}
\phi _{-}=\frac{-i}{E+mc^{2}\cos \theta }\left[ \hbar c\frac{d\phi _{+}}{dx}%
+\left( mc^{2}+V_{s}\right) \sin \theta \,\phi _{+}\right]  \label{7}
\end{equation}

\noindent one finds%
\begin{equation}
J^{1}=\frac{2\hbar c^{2}}{E+mc^{2}\cos \theta }\,\text{Im}\left( \phi
_{+}^{\ast }\frac{d\phi _{+}}{dx}\right)
\end{equation}%
Inserting (\ref{7}) into (\ref{6b}) one arrives at the following
second-order differential equation for $\phi _{+}$:
\begin{equation}
-\frac{\hbar ^{2}}{2}\frac{d^{2}\phi _{+}}{dx^{2}}+V_{\mathtt{eff}}\,\phi
_{+}=E_{\mathtt{eff}}\,\phi _{+}  \label{8}
\end{equation}%
where%
\begin{equation}
V_{\mathtt{eff}}=\frac{\sin ^{2}\theta }{2c^{2}}V_{s}^{2}+\frac{mc^{2}+E\cos
\theta }{c^{2}}V_{s}-\frac{\hbar \sin \theta }{2c}\frac{dV_{s}}{dx}
\label{v}
\end{equation}%
and%
\begin{equation}
E_{\mathtt{eff}}=\frac{E^{2}-m^{2}c^{4}}{2c^{2}}  \label{e}
\end{equation}

\noindent Therefore, the solution of the relativistic problem for this class
is mapped into a Sturm-Liouville problem for the upper component of the
Dirac spinor. In this way one can solve the Dirac problem for determining
the possible discrete or continuous eigenvalues of the system by recurring
to the solution of a Schr\"{o}dinger-like problem. For the case of a pure
scalar coupling ($E\neq 0$), it is also possible to write a second-order
differential equation for $\phi _{-}$ just differing from the equation for $%
\phi _{+}$ in the sign of the term involving $dV_{s}/dx$, namely,
\begin{equation}
-\frac{\hbar ^{2}}{2}\frac{d^{2}\phi _{\pm }}{dx^{2}}+\left( \frac{V_{s}^{2}%
}{2c^{2}}+mV_{s}\mp \frac{\hbar }{2c}\frac{dV_{s}}{dx}\right) \phi _{\pm
}=E_{\mathtt{eff}}\,\phi _{\pm }  \label{esc}
\end{equation}

\noindent This supersymmetric structure of the two-dimensional Dirac
equation with a pure scalar coupling has already been appreciated in the
literature \cite{coo}.

\section{The smooth step potential}

Now the scalar potential takes the form%
\begin{equation}
V_{s}=v_{0}\tanh \gamma x
\end{equation}%
where the skew positive parameter $\gamma $ is related to the range of the
interaction which makes $V_{s}$ to change noticeably in the interval $%
-1/\gamma <x<1/\gamma $, and $v_{0}$ is the height of the potential at $%
x=+\infty $. When $1/\gamma \gg \lambda _{C}$, where $\lambda _{C}=\hbar /mc$
is the Compton wavelength of the fermion, the potential changes smoothly
over a large distance compared to the Compton wavelength so that we can
expect the absence of quantum effects. Typical quantum effects appear when $%
1/\gamma $ is comparable to the Compton wavelength, and relativistic quantum
effects are expected when $1/\gamma $ is of the same order or smaller than
the Compton wavelength. Notice that as $\gamma \rightarrow \infty $, the
case of an extreme relativistic regime, the smooth step approximates the
sign potential already considered in Ref. \cite{aop}.

Our problem is to solve the set of equations (\ref{6a})-(\ref{6b}) for $\phi
$ and to determine the allowed energies for both classes of solutions
sketched in Sec. 2.

\subsection{The case $E=-mc^{2}\cos \protect\theta $}

As commented before, there is no solution for $\sin \theta =0$, and the
normalizable solution for $\sin \theta \neq 0$ requires $|v_{0}|>mc^{2}$:%
\begin{equation}
\phi =\left(
\begin{array}{c}
1 \\
i\cot \theta
\end{array}%
\right) N_{>}\,f  \label{s1}
\end{equation}%
for $v_{0}>mc^{2}$, and%
\begin{equation}
\phi =\left(
\begin{array}{c}
0 \\
1%
\end{array}%
\right) N_{<}\,f  \label{s2}
\end{equation}%
for $v_{0}<-mc^{2}$. Here,
\begin{equation}
f=\frac{\exp \left( -\alpha _{1}x\right) }{\cosh ^{\alpha _{2}}\gamma x}
\end{equation}%
where%
\begin{equation}
\alpha _{1}=\frac{\text{sgn}\left( v_{0}\right) mc\sin \theta }{\hbar }%
,\quad \alpha _{2}=\frac{|v_{0}|\sin \theta }{\hbar c\gamma }
\end{equation}%
The normalization condition $\int_{-\infty }^{+\infty }dx\,\left( |\phi
_{+}|^{2}+|\phi _{-}|^{2}\right) =1$ and (\ref{A1}) allow one to determine $%
N_{\gtrless }$. In the way indicated we found
\begin{equation}
N_{>}=N_{<}\sin \theta =\frac{\sin \theta }{2^{\alpha _{2}}}\sqrt{\frac{%
2\gamma }{B\left( \alpha _{+},\alpha _{-}\right) }}  \label{ene}
\end{equation}%
where%
\begin{equation}
\alpha _{\pm }=\alpha _{2}\pm \frac{\alpha _{1}}{\gamma }
\end{equation}%
From (\ref{s1}) and (\ref{s2}), one readily finds the position probability
density to be%
\begin{equation}
|\phi |^{2}=\frac{2\gamma f^{2}}{2^{2\alpha _{2}}B\left( \alpha _{+},\alpha
_{-}\right) }  \label{den}
\end{equation}%
Therefore, a massive fermion tends to concentrate at the left (right) region
when $v_{0}>0$ ($v_{0}<0$), and tends to avoid the origin more and more as $%
\sin \theta $ decreases. A massless fermion has a position probability
density symmetric around the origin. One can see that the best localization
occurs for a pure scalar coupling. In fact, the fermion becomes delocalized
as $\sin \theta $ decreases. From (\ref{A8a}) and%
\begin{equation}
\lim_{\gamma \rightarrow \infty }f=\exp \left\{ -\frac{\sin \theta }{\hbar c}%
\left[ |v_{0}|+mc^{2}\text{sgn}\left( v_{0}x\right) \right] |x|\right\}
\end{equation}%
one recovers the value for $\phi $ in the case of the sign potential (at
large $\gamma $) as in Ref. \cite{aop}. Figure 1 illustrates the position
probability density for a massive fermion with $v_{0}/mc^{2}=2$, $\theta
=3\pi /8$ and two different values of $\gamma $. From this figure one sees
that $|\phi |^{2}$ shrinks with rising $\gamma $.

The expectation value of $x$ and $x^{2}$ is given by%
\begin{equation}
<x>=-\frac{4\gamma }{2^{2\alpha _{2}}B\left( \alpha _{+},\alpha _{-}\right) }%
\int_{0}^{\infty }dx\,\frac{x\sinh 2\alpha _{1}x}{\cosh ^{2\alpha
_{2}}\gamma x}  \label{espec1}
\end{equation}%
and%
\begin{equation}
<x^{2}>=\frac{4\gamma }{2^{2\alpha _{2}}B\left( \alpha _{+},\alpha
_{-}\right) }\int_{0}^{\infty }dx\,\frac{x^{2}\cosh 2\alpha _{1}x}{\cosh
^{2\alpha _{2}}\gamma x}  \label{espec2}
\end{equation}%
From (\ref{A7a}) and (\ref{A7b}) these last results can be simplified to%
\begin{equation}
<x>=-\frac{1}{2\gamma }{\Delta }\left( \alpha \right)
\end{equation}%
and%
\begin{equation}
<x^{2}>=\frac{1}{\left( 2\gamma \right) ^{2}}\,{\Sigma }^{\left( 1\right)
}\left( \alpha \right) +<x>^{2}
\end{equation}%
and hence the fermion is confined within an interval $\Delta x=\sqrt{%
<x^{2}>-<x>^{2}}$ given by%
\begin{equation}
\Delta x=\frac{1}{2\gamma }\sqrt{{\Sigma }^{\left( 1\right) }\left( \alpha
\right) }  \label{dx}
\end{equation}%
Thereby, with the help of (\ref{A8b}), one obtains the values for $<x>$ and $%
\Delta x$ either in the case of $\sin \theta \rightarrow 0$ or in the case
of the sign potential (at large $\gamma $) as in Ref. \cite{aop}:
\begin{subequations}
\begin{eqnarray}
&<&x>\rightarrow -\text{sgn}\left( v_{0}\right) \frac{\hbar c}{\sin \theta }%
\frac{mc^{2}}{v_{0}^{2}-m^{2}c^{4}}  \label{g1a} \\
&&  \notag \\
\Delta x &\rightarrow &\frac{\hbar c}{\sqrt{2}\sin \theta }\frac{\sqrt{%
v_{0}^{2}+m^{2}c^{4}}}{v_{0}^{2}-m^{2}c^{4}}  \label{g1b}
\end{eqnarray}%
On the other hand, from (\ref{A9}) one sees that when $\gamma \rightarrow 0$
or $|v_{0}|\rightarrow \infty $

\end{subequations}
\begin{subequations}
\begin{eqnarray}
&<&x>\rightarrow \frac{1}{2\gamma }\ln \frac{|v_{0}|-\text{sgn}\left(
v_{0}\right) mc^{2}}{|v_{0}|+\text{sgn}\left( v_{0}\right) mc^{2}}
\label{g2a} \\
&&  \notag \\
\Delta x &\rightarrow &\sqrt{\frac{\hbar c}{2\gamma \sin \theta }\frac{%
|v_{0}|}{v_{0}^{2}-m^{2}c^{4}}}  \label{g2b}
\end{eqnarray}%
Again one can see that the fermion becomes delocalized as $\sin \theta $
decreases and that the best localization occurs for a pure scalar coupling.
More than this, $<x>\rightarrow -\infty $ and $\Delta x\rightarrow \infty $
as $|v_{0}|\rightarrow mc^{2}$, and besides $<x>\rightarrow 0$ and $\Delta
x\rightarrow 0$ as $|v_{0}|\rightarrow \infty $.

If $\Delta x$ reduces its extension (with rising $|v_{0}|$ or $\sin \theta $
or $\gamma $) then $\Delta p$ must expand, in consonance with the Heisenberg
uncertainty principle. Nevertheless, the maximum uncertainty in the momentum
is comparable with $mc$ requiring that is impossible to localize a fermion
in a region of space less than or comparable with half of its Compton
wavelength (see, for example, \cite{gre}). This impasse can be broken by
resorting to the concepts of effective mass and effective Compton
wavelength. Indeed, if one defines an effective mass as $m_{\mathtt{eff}}=m%
\sqrt{1+\left( v_{0}/mc^{2}\right) ^{2}}$ and an effective Compton
wavelength $\lambda _{\mathtt{eff}}=\hbar /\left( m_{\mathtt{eff}}c\right) $%
, one will find
\end{subequations}
\begin{equation}
\Delta x=\frac{\sqrt{2}\lambda _{\mathtt{eff}}}{4\sin \theta }\sqrt{\left(
\alpha _{+}^{2}+\alpha _{-}^{2}\right) {\Sigma }^{\left( 1\right) }\left(
\alpha \right) }  \label{dx1}
\end{equation}%
It follows that the high localization of fermions, related to high values of
$|v_{0}|$ and $\gamma $, never menaces the single-particle interpretation of
the Dirac theory even if the fermion is massless ($m_{\mathtt{eff}%
}=|v_{0}|/c^{2}$). This fact is convincing because the scalar coupling
exceeds the vector coupling, and so the conditions for Klein's paradox are
never reached. As a matter of fact, (\ref{g1b}) furnishes $\left( \Delta
x\right) _{\min }\simeq \lambda _{\mathtt{eff}}/(\sqrt{2}\sin \theta )$ for $%
|v_{0}|\gg mc^{2}$ and $\hbar \gamma \gg mc$.

\subsection{The case $E\neq -mc^{2}\cos \protect\theta $}

For our model, recalling (\ref{7}) and (\ref{v}), one finds%
\begin{equation}
\phi _{-}=\frac{-i}{E+mc^{2}\cos \theta }\left[ \hbar c\frac{d\phi _{+}}{dx}%
\,+\left( mc^{2}+v_{0}\,\tanh \gamma x\right) \sin \theta \,\phi _{+}\right]
\end{equation}%
and%
\begin{equation}
V_{\mathtt{eff}}=-V_{1}\text{\textrm{sech}}^{2}\gamma x+V_{2}\tanh \gamma
x+V_{3}  \label{pot}
\end{equation}%
where the following abbreviations have been used:

\begin{subequations}
\begin{eqnarray}
V_{1} &=&v_{0}\sin \theta \,\frac{v_{0}\sin \theta +\hbar c\gamma }{2c^{2}}
\label{par1} \\
&&  \notag \\
V_{2} &=&v_{0}\,\frac{E\cos \theta +mc^{2}}{c^{2}}  \label{par2} \\
&&  \notag \\
V_{3} &=&\frac{v_{0}^{2}\sin ^{2}\theta }{2c^{2}}  \label{par3}
\end{eqnarray}%
It is instructive to note that if we let $\gamma \rightarrow \infty $, then $%
\tanh \gamma x\rightarrow $ sgn$\left( x\right) $ and $\left( \gamma
/2\right) $\textrm{sech}$^{2}\gamma x\rightarrow \delta \left( x\right) $.
For $\sin \theta =0$, the \textquotedblleft effective
potential\textquotedblright\ is an ascendant (a descendant) smooth step if $%
V_{2}>0$ ($V_{2}<0$). For $\sin \theta \neq 0$, though, the
\textquotedblleft effective potential\textquotedblright\ has the same form
as the exactly solvable Rosen-Morse potential \cite{ros}, \cite{nie}. The
Rosen-Morse potential approaches $V_{3}\pm V_{2}$ as $x\rightarrow \pm
\infty $ and has an extremum when $|V_{2}|<2|V_{1}|$ at

\end{subequations}
\begin{equation}
x_{m}=\frac{1}{2\gamma }\ln \left( \frac{2V_{1}-V_{2}}{2V_{1}+V_{2}}\right)
\label{xm}
\end{equation}

\noindent As a matter of fact, potential-well structures can be achieved
when $|V_{2}|<2|V_{1}|$ $\ $with $V_{1}>0$. To acknowledge that the
effective potential for the mixing given by (\ref{5}) is a Rosen-Morse
potential can help you to see more clearly how a kink-like smooth step
potential might furnish a finite set of bound-state solutions. After all, we
shall not use the knowledge about the exact analytical solution for the
Rosen-Morse potential.

\subsubsection{The asymptotic solutions}

As $|x|\gg 1/\gamma $ the effective potential is practically constant (the
main transition region occurs in $|x|<1/\gamma $) and the solutions for the
Dirac equation can be approximate by those ones for a free particle.
Furthermore, the asymptotic behaviour will show itself suitable to impose
the appropriate boundary conditions to the complete solution to the problem.

We turn our attention to scattering states for fermions coming from the
left. Then, $\phi $ for $x\rightarrow -\infty $ describes an incident wave
moving to the right and a reflected wave moving to the left, and $\phi $ for
$x\rightarrow +\infty $ describes a transmitted wave moving to the right or
an evanescent wave. The upper components for scattering states are written
as
\begin{equation}
\phi _{+}=\left\{
\begin{array}{cc}
A_{+}e^{+ik_{-}x}+A_{-}e^{-ik_{-}x}, & \text{for\quad }x\rightarrow -\infty
\\
&  \\
B_{\pm }e^{\pm ik_{+}x}, & \text{for\quad }x\rightarrow +\infty%
\end{array}%
\right.  \label{phi}
\end{equation}%
where

\begin{equation}
\hbar k_{\pm }=\sqrt{2\left( E_{\mathtt{eff}}-V_{3}\mp V_{2}\right) }
\end{equation}%
Note that $k_{+}$ is a real number for a progressive wave and an imaginary
number for an evanescent wave ($k_{-}$ is a real number for scattering
states). Therefore,%
\begin{equation}
J^{1}\left( -\infty \right) =\frac{2\hbar c^{2}k_{-}}{E+mc^{2}\cos \theta }%
\left( |A_{\pm }|^{2}-|A_{\mp }|^{2}\right) ,\text{\quad for\quad }E\gtrless
-mc^{2}\cos \theta
\end{equation}%
and
\begin{equation}
J^{1}\left( +\infty \right) =\pm \,\frac{2\hbar c^{2}\text{Re}\,k_{+}}{%
E+mc^{2}\cos \theta }\,|B_{\pm }|^{2},\text{\quad for\quad }E\gtrless
-mc^{2}\cos \theta
\end{equation}%
Note that $J^{1}\left( -\infty \right) =J_{\mathtt{inc}}-J_{\mathtt{ref}}$
and $J^{1}\left( +\infty \right) =J_{\mathtt{tran}}$, where $J_{\mathtt{inc}}
$, $J_{\mathtt{ref}}$ and $J_{\mathtt{tran}}$ are nonnegative quantities
characterizing the incident, reflected and transmitted waves, respectively.
Note also that the roles of $A_{+}$ and $A_{-}$ are exchanged as the sign of
$E+mc^{2}\cos \theta $ changes. In fact, if $E>-mc^{2}\cos \theta $, then $%
A_{+}e^{+ik_{-}x}$ ($A_{-}e^{-ik_{-}x}$) will describe the incident
(reflected) wave, and $B_{-}=0$. On the other hand, if $E<-mc^{2}\cos \theta
$, then $A_{-}e^{-ik_{-}x}$ ($A_{+}e^{+ik_{-}x}$) will describe the incident
(reflected) wave, and $B_{+}=0$. Therefore, the reflection and transmission
amplitudes are given by%
\begin{equation}
r=\frac{A_{\mp }}{A_{\pm }},\text{\quad }t=\frac{B_{\pm }}{A_{\pm }},\text{%
\quad for\quad }E\gtrless -mc^{2}\cos \theta   \label{t}
\end{equation}%
To determine the transmission coefficient we use the current densities $%
J^{1}\left( -\infty \right) $ and $J^{1}\left( +\infty \right) $. The $x$%
-independent space component of the current allows us to define the
reflection and transmission coefficients as%
\begin{equation}
R=\frac{|A_{\mp }|^{2}}{|A_{\pm }|^{2}},\text{\quad }T=\frac{\text{Re}\,k_{+}%
}{k_{-}}\frac{|B_{\pm }|^{2}}{|A_{\pm }|^{2}},\text{\quad for\quad }%
E\gtrless -mc^{2}\cos \theta   \label{tr}
\end{equation}%
Notice that $R+T=1$ by construction.

\subsubsection{The complete solutions}

Armed with the knowledge about asymptotic solutions and with the definition
of the transmission coefficient we proceed for searching solutions on the
entire region of space.

Changing the independent variable $x$ in (\ref{8}) to

\begin{equation}
y=\frac{1}{2}\left( 1-\tanh \gamma x\right)  \label{xtoy}
\end{equation}%
the differential equation is transformed into

\begin{equation}
y\left( 1-y\right) \frac{d^{2}\phi _{+}}{dy^{2}}+\left( 1-2y\right) \frac{%
d\phi _{+}}{dy}+\Theta \phi _{+}=0  \label{eq11}
\end{equation}

\noindent where

\begin{equation}
\Theta =\frac{4V_{1}y\left( 1-y\right) -V_{2}\left( 1-2y\right) -V_{3}+E_{%
\mathtt{eff}}}{2\left( \hbar \gamma \right) ^{2}y\left( 1-y\right) }
\label{eq1a}
\end{equation}%
regardless of the sign of $x$. Introducing a new function $\varphi (y)$
through the relation%
\begin{equation}
\phi _{+}(y)=y^{\nu }\left( 1-y\right) ^{\mu }\varphi (y)  \label{rel}
\end{equation}

\noindent and defining

\begin{subequations}
\label{def}
\begin{eqnarray}
a &=&\mu +\nu +\frac{1-\omega }{2},\quad b=\mu +\nu +\frac{1+\omega }{2}%
,\quad d=2\nu +1  \label{def1} \\
&&  \notag \\
\mu ^{2} &=&-\left( \frac{k_{-}}{2\gamma }\right) ^{2},\quad \nu
^{2}=-\left( \frac{k_{+}}{2\gamma }\right) ^{2},\quad \omega ^{2}=1+\frac{%
8V_{1}}{\left( \hbar \gamma \right) ^{2}}  \label{def2}
\end{eqnarray}%
Eq. (\ref{eq11}) becomes the hypergeometric differential equation \cite{abr}

\end{subequations}
\begin{equation}
y\left( 1-y\right) \frac{d^{2}\varphi }{dy^{2}}+\left[ d-\left( a+b+1\right)
y\right] \frac{d\varphi }{dy}-ab\varphi =0  \label{hyper}
\end{equation}

\noindent whose general solution can be written in terms of the Gauss
hypergeometric series

\begin{equation}
_{2}F_{1}\left( a,b,d,y\right) =\frac{\Gamma \left( d\right) }{\Gamma \left(
a\right) \Gamma \left( b\right) }\sum\limits_{n=0}^{\infty }\frac{\Gamma
\left( a+n\right) \Gamma \left( b+n\right) }{\Gamma \left( d+n\right) }\frac{%
y^{n}}{n!}  \label{Gauss hyper}
\end{equation}

\noindent in the form \cite{abr}

\begin{equation}
\varphi =A\;_{2}F_{1}\left( a,b,d,y\right) +By^{-2\nu }\;_{2}F_{1}\left(
a+1-d,b+1-d,2-d,y\right)  \label{gen1}
\end{equation}

\noindent in such a way that

\begin{equation*}
\phi _{+}=A\,y^{\nu }\left( 1-y\right) ^{\mu }\;_{2}F_{1}\left(
a,b,d,y\right)
\end{equation*}

\begin{equation}
+\,B\,y^{-\nu }\left( 1-y\right) ^{\mu }\;_{2}F_{1}\left(
a+1-d,b+1-d,2-d,y\right)  \label{phi1}
\end{equation}

\noindent with the constants $A$ and $B$ to be fitted by the asymptotic
behaviour analyzed in the previous discussion.

As $x\rightarrow +\infty $ (that is, as $y\rightarrow 0$), one has that $%
y\simeq \exp \left( -2\gamma x\right) $ and (\ref{phi1}), because $%
_{2}F_{1}\left( a,b,d,0\right) =1$, reduces to

\begin{equation}
\phi _{+}\left( +\infty \right) \simeq Ae^{-2\gamma \nu x}+Be^{2\gamma \nu x}
\label{phi9}
\end{equation}

\noindent so the asymptotic behaviour, for $\nu =\mp ik_{+}/\left( 2\gamma
\right) $, requires that $B=0$ and $A=B_{\pm }$ corresponding to $E\gtrless
-mc^{2}\cos \theta $, or equivalently $A=0$ and $B=B_{\mp }$ corresponding
to $E\lessgtr -mc^{2}\cos \theta $. We choose $B=0$.

The asymptotic behaviour as $x\rightarrow -\infty $ ($y\rightarrow 1$) can
be found by using the relation for passing over from $y$ to $1-y$:

\begin{equation*}
_{2}F_{1}\left( a,b,d,y\right) =\gamma _{-}\;_{2}F_{1}\left(
a,b,a+b-d+1,1-y\right)
\end{equation*}

\begin{equation}
+\,\,\gamma _{+}\;_{2}F_{1}\left( d-a,d-b,d-a-b+1,1-y\right) \left(
1-y\right) ^{d-a-b}  \label{cona}
\end{equation}%
where $\gamma _{+}$ and $\gamma _{-}$ are expressed in terms of the gamma
function as

\begin{equation}
\gamma _{-}=\frac{\Gamma \left( d\right) \Gamma \left( d-a-b\right) }{\Gamma
\left( d-a\right) \Gamma \left( d-b\right) },\quad \gamma _{+}=\frac{\Gamma
\left( d\right) \Gamma \left( a+b-d\right) }{\Gamma \left( a\right) \Gamma
\left( b\right) }\quad  \label{g1}
\end{equation}

\noindent which can also be written as

\begin{equation}
\gamma _{\pm }=\frac{\Gamma \left( 2\nu +1\right) \Gamma \left( \pm 2\mu
\right) }{\Gamma \left( \frac{1+\omega }{2}+\nu \pm \mu \right) \Gamma
\left( \frac{1-\omega }{2}+\nu \pm \mu \right) }
\end{equation}%
Now, as $x\rightarrow -\infty $, $1-y\simeq \exp \left( +2\gamma x\right) $.
This time, (\ref{phi1}) tends to

\begin{equation}
\phi _{+}\left( -\infty \right) \simeq A\gamma _{+}e^{-2\gamma \mu
x}+A\gamma _{-}e^{+2\gamma \mu x}  \label{phi10}
\end{equation}

\noindent so that $A\gamma _{+}=A_{\pm }$ and $A\gamma _{-}=A_{\mp }$ for $%
\mu =\mp ik_{-}/\left( 2\gamma \right) $, in accordance with the previous
analysis for very large negative values of $x$.

Those asymptotic behaviours are all one needs to determinate the
transmission amplitude (\ref{t}) and the transmission coefficient (\ref{tr}%
). Now, these quantities can now be expressed in terms of $\gamma _{\pm }$ as%
\begin{equation}
\text{\quad }t=\frac{1}{\gamma _{\pm }},\text{\quad }T=\left\vert \frac{%
\text{Im}\nu }{\mu }\right\vert \frac{1}{|\gamma _{\pm }|^{2}},\text{\quad
for\quad }\mu =\mp \frac{ik_{-}}{2\gamma }  \label{t2}
\end{equation}%
Notice that $\omega ^{2}$ in (\ref{def2}) can be written as%
\begin{equation}
\omega ^{2}=\left( 1+\frac{2v_{0}\sin \theta }{\hbar c\gamma }\right) ^{2}
\end{equation}%
so that%
\begin{equation}
\omega =\pm \left( 1+\frac{2v_{0}\sin \theta }{\hbar c\gamma }\right)
\end{equation}%
Furthermore, by using the following identities \cite{abr}

\begin{equation}
|\Gamma \left( iv\right) |^{2}=\frac{\pi }{v\sinh \pi v},\quad |\Gamma
\left( 1+iv\right) |^{2}=\frac{\pi v}{\sinh \pi v}  \label{gama3}
\end{equation}%
added by the identity \cite{mgg}%
\begin{equation}
|\Gamma \left( u+iv\right) \Gamma \left( 1-u+iv\right) |^{2}=\frac{2\pi ^{2}%
}{\cosh 2\pi v-\cos 2\pi u}  \label{gama2}
\end{equation}

\noindent where $u$ and $v$ are the real and imaginary parts of a complex
number, one can show that

\begin{equation}
T=\frac{2\sinh \frac{k_{-}\pi }{\gamma }\sinh \text{Re}\,\frac{k_{+}\pi }{%
\gamma }}{\left\vert \cosh \frac{\left( k_{-}+k_{+}\right) \pi }{\gamma }%
+\cos \pi \omega \right\vert }  \label{t3}
\end{equation}

\noindent taking no regard if $E>-mc^{2}\cos \theta $ or $E<-mc^{2}\cos
\theta $. Nevertheless, scattering states are possible only if $|E+v_{0}\cos
\theta |>|mc^{2}-v_{0}|$ because $k_{-}$ is a real number, and there is a
transmitted wave only if $|E-v_{0}\cos \theta |>|mc^{2}+v_{0}|$. As $%
|E|\rightarrow \infty $, $T\rightarrow 1$ as it should be. Seen as a
function of $E$, for $E>-mc^{2}\cos \theta $, the transmission coefficient
presents a profile typical for the nonrelativistic scattering in a step
potential. Seen as a function of the mixing angle the transmission
coefficient presents some intriguing results explained by observing that the
effective potential presents an ascendant (descendant) step for small
(large) values of $\theta $ (see Ref. \cite{aop} for $\gamma \rightarrow
\infty $). The transmission coefficient vanishes for enough small mixing
angles and energies because the effective energy is smaller than the height
of the effective step potential. For $|v_{0}|>mc^{2}$, the absence of
scattering for enough large mixing angles and enough small energies occurs
because the effective energy is smaller than the effective step potential in
the region of incidence.

By the way, as $\gamma \rightarrow 0$ one finds%
\begin{equation}
T=\left\{
\begin{array}{cc}
1, & \text{for\quad }k_{+}\in
\mathbb{R}
\\
&  \\
0, & \text{for\quad }k_{+}=\pm i|k_{+}|%
\end{array}%
\right.
\end{equation}%
reflecting our expectation about the absence of quantum effects for a
potential whose interval of appreciable variation is much more larger than
the Compton wavelength. It is remarkable that this \textquotedblleft
classical\textquotedblright\ scattering also takes place for massless
fermions. On the other hand, for $\gamma \rightarrow \infty $ one finds the
transmission coefficient for the sign step potential \cite{aop}:

\begin{equation}
T\simeq \frac{4k_{-}\text{Re}\,k_{+}}{\left( k_{-}+k_{+}\right) ^{2}+\left(
\frac{2v_{0}\sin \theta }{_{\hbar c}}\right) ^{2}}  \label{16}
\end{equation}

\noindent as it should be.

\subsubsection{Bound states}

The possibility of bound states requires a solution with an asymptotic
behaviour given by (\ref{phi}) with $k_{\pm }=i|k_{\pm }|$ and $%
A_{+}=B_{-}=0 $, or $k_{\pm }=-i|k_{\pm }|$ and $A_{-}=B_{+}=0$, to obtain a
square-integrable $\phi _{+}$, meaning that%
\begin{equation}
E_{\mathtt{eff}}<V_{3}\pm V_{2}  \label{ceff}
\end{equation}%
or equivalently%
\begin{equation}
|E\pm v_{0}\cos \theta |<|mc^{2}\mp v_{0}|
\end{equation}%
On the other hand, if one considers the transmission amplitude $t$ in (\ref%
{t2}) as a function of the complex variables $k_{\pm }$ one sees that for $%
k_{\pm }>0$ ($\mu $ and $\nu $ as imaginary quantities) one obtains the
scattering states whereas the bound states would be obtained by the poles
lying along the imaginary axis of the complex $k$-plane. From (\ref{phi9})
with $B=0$ one sees that $\nu $ is a positive quantity. On the other hand, $%
\mu $ is positive (negative) if $\gamma _{+}=0$ ($\gamma _{-}=0$). The poles
of the transmission amplitude are given by the zeros of $\gamma _{\pm }$. It
happens that $\Gamma \left( z\right) $ has no zeros but it has simple poles
on the real axis at $z=-n$ with $n=0,1,2,\ldots $ Because $_{2}F_{1}\left(
a,b,d,y\right) $ is invariant under exchange of $a$ and $b$, the
quantization condition is thus given by $a=-n$ or $b=-n$. Therefore, the
bound states occur only for%
\begin{equation}
|\mu |+|\nu |+\frac{1-|\omega |}{2}=-n  \label{cq0}
\end{equation}%
Recalling the definitions of $\mu $, $\nu $ and $\omega $ given in (\ref{def}%
), the quantization condition can be rewritten as%
\begin{equation*}
\sqrt{\left( mc^{2}+v_{0}\right) ^{2}-\left( E-v_{0}\cos \theta \right) ^{2}}%
+\sqrt{\left( mc^{2}-v_{0}\right) ^{2}-\left( E+v_{0}\cos \theta \right) ^{2}%
}
\end{equation*}

\begin{equation}
=2\hbar c\gamma \left( \pm \frac{v_{0}\sin \theta }{\hbar c\gamma }-N\right)
,\text{\quad for\quad }\frac{v_{0}\sin \theta }{\hbar c\gamma }\gtrless -%
\frac{1}{2}  \label{cq1}
\end{equation}%
with%
\begin{equation}
N=\left\{
\begin{array}{c}
n, \\
\\
n+1,%
\end{array}%
\begin{array}{c}
\text{\quad for\quad }\frac{v_{0}\sin \theta }{\hbar c\gamma }>-\frac{1}{2}
\\
\\
\text{\quad for\quad }\frac{v_{0}\sin \theta }{\hbar c\gamma }<-\frac{1}{2}%
\end{array}%
\right.
\end{equation}%
Because the first line of (\ref{cq1}) is a positive number and $n$ is a
nonnegative integer, one finds supplementary restrictions imposed on $%
v_{0}\sin \theta /\left( \hbar c\gamma \right) $ and $N$:
\begin{equation}
\frac{v_{0}\sin \theta }{\hbar c\gamma }>0,\text{\quad }N<\frac{v_{0}\sin
\theta }{\hbar c\gamma }  \label{cv1a}
\end{equation}%
or%
\begin{equation}
\frac{v_{0}\sin \theta }{\hbar c\gamma }<-1,\text{\quad }N<-\frac{v_{0}\sin
\theta }{\hbar c\gamma }  \label{cv1b}
\end{equation}%
This means that there is a finite set of bound-state solutions depending on
the sign and size of $v_{0}\sin \theta /\left( \hbar c\gamma \right) $, and
that the number of allowed solutions increases with $|v_{0}|\sin \theta
/\left( \hbar c\gamma \right) $. It is worth to mention that the threshold ($%
|v_{0}|_{\min }$) is an increasing monotonic function of $\gamma $ with $%
|v_{0}|_{\min }\rightarrow mc^{2}$ as $\gamma \rightarrow 0$ so that the
existence of those bound-state solutions is not workable in a
nonrelativistic scheme. In particular, there is no bound-state solution
neither when $\sin \theta =0$ nor in the limiting case $\gamma \rightarrow
\infty $. Furthermore, the symmetries related to the charge-conjugation and
chiral-conjugation operations discussed in Sec. 2 are clearly revealed. It
is interesting to remark that the conditions on $v_{0}\sin \theta /\left(
\hbar c\gamma \right) $ in (\ref{cv1a}) and (\ref{cv1b}) are these ones that
make $V_{1}>0$ in (\ref{par1}). \

The irrational equation (\ref{cq1}) can be solved iteratively to determine
the eigenenergies. However, if one squares (\ref{cq1}), the resulting
quantization condition can also be expressed as a second-order algebraic
equation in $E$ with two branches of solutions%
\begin{equation}
E=\frac{-c_{1}\pm \sqrt{c_{1}^{2}-4c_{2}c_{0}}}{2c_{2}}  \label{alg1}
\end{equation}%
where
\begin{subequations}
\label{alg2}
\begin{eqnarray}
c_{2} &=&v_{0}^{2}+(\hbar c\gamma N)^{2}-2\left( \hbar c\gamma N\right)
|v_{0}|\sin \theta  \label{alg2a} \\
&&  \notag \\
c_{1} &=&2mc^{2}v_{0}^{2}\cos \theta  \label{alg2b} \\
&&  \notag \\
c_{0} &=&(\hbar c\gamma N)^{4}-4(\hbar c\gamma N)^{3}|v_{0}|\sin \theta
+(\hbar c\gamma N)^{2}\left( 5v_{0}^{2}\sin ^{2}\theta -m^{2}c^{4}\right)
\notag \\
&&  \notag \\
&&+2(\hbar c\gamma N)\left( m^{2}c^{4}-v_{0}^{2}\sin ^{2}\theta \right)
|v_{0}|\sin \theta +m^{2}c^{4}v_{0}^{2}\cos ^{2}\theta  \label{alg2c}
\end{eqnarray}%
The price paid by those analytical solutions is that some of them can be
spurious. Of course, the false roots can be eliminated by inspecting whether
they satisfy the original equation. Furthermore, despite the closed form for
the Dirac eigenenergies, the solutions given by (\ref{alg1}) present an
intricate dependence on $|v_{0}|$, $\gamma $, $\theta $, $m$ and $N$. There
is an evident problem with the $n=0$ solution when $v_{0}>0$ because Eq. (%
\ref{alg1}) with $n=0$ and $v_{0}>0$ presents the unique root $E=-mc^{2}\cos
\theta $. This, of course, is not a proper solution of the problem. The
results for $v_{0}<0$ are the same as those ones for $v_{0}>0$ if one
changes $|v_{0}|\sin \theta /\left( \hbar c\gamma \right) $ by $|v_{0}|\sin
\theta /\left( \hbar c\gamma \right) +1$ and $n$ by $n-1$.

Numerical solutions for the eigenenergies corresponding to the three lowest
quantum numbers ($n=1,2,3$ for $v_{0}>0$) are shown in Figures 2, 3 and 4
for a massive fermion. In all of these figures, the innermost curves
correspond to the lowest quantum numbers and the dotted line corresponds to
the isolated solution ($E=-mc^{2}\cos \theta $) discussed in the previous
section.

In Fig. 2 we show the eigenenergies as a function of $|v_{0}|/mc^{2}$ for $%
\theta =3\pi /8$ and $\hbar \gamma /mc=1/10$. Notice that a minimum value
for $|v_{0}|$ is required to obtain at least one energy level and that
because $\theta <\pi /2$ the branch of solutions with $E<-mc^{2}\cos \theta $
is more favoured. Notice also that more and more energy levels arise for
each branch as $v_{0}$ increases.

In Fig. 3 the eigenenergies are shown as a function of $\cos \theta $ for $%
v_{0}/mc^{2}=2$ and $\hbar \gamma /mc=1/10$. It is remarkable that the
eigenenergy changes from $E$ to $-E$ when $\theta $ changes from $\pi
/2-\varepsilon $ to $\pi /2+\varepsilon $. In particular, the energy levels
exhibit symmetry about $E=0$ when $\theta =\pi /2$. All of the levels tend
to vanish as $\cos \theta $ tends to $\pm 1$. Incidentally, this
disappearance of energy levels is more forceful for higher values of $n$.
The branch for $E\gtrless -mc^{2}\cos \theta $ is more favored when $\theta
\lessgtr \pi /2$.

In Fig. 4 the eigenenergies are shown as a function of $\hbar \gamma /mc$
for $\theta =3\pi /8$ and $v_{0}/mc^{2}=2$. For $\gamma \simeq 0$ we have a
very high density of energy levels. These levels correspond to very
delocalized states due to the large extension of the interaction region. The
density of energy levels decreases with increasing $\hbar \gamma /mc$. It is
worth noting that the energy levels exist in a finite interval of $\hbar
\gamma /mc$, and so they do not exist in the extreme relativistic regime for
a finite value of $|v_{0}|/mc^{2}$. It should be mentioned, though, the
upper limit of $\hbar \gamma /mc$ increases monotonously with $|v_{0}|/mc^{2}
$. Notice also that because $\theta <\pi /2$ the branch of solutions with $%
E<-mc^{2}\cos \theta $ is more favoured.

The case of a massless fermion, as already discussed before with fulcrum on
the charge-conjugation and the chiral-conjugation operations, presents a
spectrum symmetrical about $E=0$ and seen as a function of $\theta $
exhibits an additional symmetry about $\theta =\pi /2$.

Now the Gauss hypergeometric series $_{2}F_{1}\left( a,b,d,y\right) $
reduces to nothing but a polynomial of degree $n$ in $y$ when $a$ or $b$ is
equal to $-n$: Jacobi's polynomial of index $\alpha $ and $\beta $. Indeed,
for $a=-n$ one has \cite{abr}

\end{subequations}
\begin{eqnarray}
_{2}F_{1}\left( a,b,d,y\right) &=&_{2}F_{1}\left( -n,\alpha +1+\beta
+n,\alpha +1,y\right)  \notag \\
&&  \notag \\
&=&\frac{n!}{\left( \alpha +1\right) _{n}}P_{n}^{\left( \alpha ,\beta
\right) }\left( \xi \right)  \label{jaco1}
\end{eqnarray}

\noindent where

\begin{equation}
\alpha =2|\nu |,\quad \beta =2|\mu |,\quad \xi =1-2y  \label{alfabeta}
\end{equation}

\noindent and $\left( \alpha \right) _{n}=\alpha \left( \alpha +1\right)
\left( \alpha +2\right) ...\left( \alpha +n-1\right) $ with $\left( \alpha
\right) _{0}=1$. \noindent Hence $\phi _{+}$ can be written as

\begin{equation}
\phi _{+}\left( \xi \right) =N_{n}\left( 1-\xi \right) ^{\alpha /2}\left(
1+\xi \right) ^{\beta /2}P_{n}^{\left( \alpha ,\beta \right) }\left( \xi
\right)  \label{17}
\end{equation}%
and $\phi _{-}$ assumes the form%
\begin{equation}
\phi _{-}\left( \xi \right) =\frac{-i\hbar c\gamma }{E+mc^{2}\cos \theta }%
\left[ \left( 1-\xi ^{2}\right) \frac{d\phi _{+}}{d\xi }+\frac{%
mc^{2}+v_{0}\xi }{\hbar c\gamma }\sin \theta \,\phi _{+}\right]  \label{j1}
\end{equation}%
Because Jacobi's polynomials $P_{n}^{\left( \alpha ,\beta \right) }\left(
\xi \right) $ have $n$ distinct zeros \cite{abr} $\phi _{+}$ has $n$ nodes,
and this fact causes $|\phi |^{2}$ to have between $n+1$ and $2n+1$ humps.
The position probability density has a lonely hump exclusively for the
isolated solution. The determination of the normalization constant $N_{n}$
looks exceedingly complicated and so we content ourselves with a numerical
illustration. Figure 5 shows the normalized position probability density for
a massive fermion for the Sturm-Liouville solution with $n=1$, $%
v_{0}/mc^{2}=2$, $\hbar \gamma /mc=1/10$ and $\theta =3\pi /8$.

\section{Final remarks}

We have assessed the stationary states of a fermion under the influence of
the kink-like potential $\tanh \gamma x$ as a generalization of the sign
potential (see \cite{aop}). Several interesting properties arose depending
on the size of the skew parameter $\gamma $. For a special mixing of scalar
and vector couplings, a continuous chiral-conjugation transformation was
allowed to decouple the upper and lower components of the Dirac spinor and
to assess the scattering problem under a Sturm-Liouville perspective. A
finite set of intrinsically relativistic bound-state solutions was computed
directly from the poles of the transmission amplitude. An isolated solution
from the Sturm-Liouville problem corresponding to a bound state was also
analyzed. The concepts of effective mass and effective Compton wavelength
were used to show the impossibility of pair production under a strong
potential despite the high localization of the fermion. It was also shown
that all of the bound-state solutions disappear asymptotically as one
approaches the conditions for the realization of \textquotedblleft spin and
pseudospin symmetries\textquotedblright .

\appendix

\section{Useful integrals and limits}

The integral necessary for calculating the normalization constants in (\ref%
{s1}) and (\ref{s2}) is tabulated (see the formula 3.512.1, or 8.380.10, in
Ref. \cite{gr}):%
\begin{equation}
\int_{0}^{\infty }dx\,\frac{\cosh 2\beta _{1}x}{\cosh ^{2\beta _{2}}\gamma x}%
=\frac{2^{2\beta _{2}}}{4\gamma }B\left( \beta _{2}+\frac{\beta _{1}}{\gamma
},\beta _{2}-\frac{\beta _{1}}{\gamma }\right)  \tag{A1}  \label{A1}
\end{equation}%
where%
\begin{equation}
B\left( z_{1},z_{2}\right) =\frac{\Gamma \left( z_{1}\right) \Gamma \left(
z_{2}\right) }{\Gamma \left( z_{1}+z_{2}\right) },\text{\quad Re\thinspace }%
z_{1}>0,\text{\quad Re\thinspace }z_{2}>0  \tag{A2}
\end{equation}%
is the beta function \cite{abr}.

We now proceed to evaluate the integrals in (\ref{espec1}) and (\ref{espec2}%
). We introduce an accessory parameter $\lambda $ in such a way that%
\begin{equation}
\int_{0}^{\infty }dx\,\frac{x\sinh 2\lambda \beta _{1}x}{\cosh ^{2\beta
_{2}}\gamma x}=\frac{1}{2\beta _{1}}\frac{\partial I\left( \lambda \right) }{%
\partial \lambda }  \tag{A3a}
\end{equation}%
\begin{equation}
\int_{0}^{\infty }dx\,\frac{x^{2}\cosh 2\lambda \beta _{1}x}{\cosh ^{2\beta
_{2}}\gamma x}=\frac{1}{4\beta _{1}^{2}}\frac{\partial ^{2}I\left( \lambda
\right) }{\partial \lambda ^{2}}  \tag{A3b}
\end{equation}%
where$^{{}}$%
\begin{equation}
I\left( \lambda \right) =\int_{0}^{\infty }dx\,\frac{\cosh 2\lambda \beta
_{1}x}{\cosh ^{2\beta _{2}}\gamma x}=\frac{2^{2\beta _{2}}}{4\gamma }B\left(
\beta _{+},\beta _{-}\right)   \tag{A4}
\end{equation}%
and $\beta _{\pm }\left( \lambda \right) =\beta _{2}\pm \lambda \beta
_{1}/\gamma $. Defining%
\begin{equation}
{\Delta }\left( \beta \right) =\boldsymbol{\psi \,}\left( \beta _{+}\right) -%
\boldsymbol{\psi \,}\left( \beta _{-}\right) ,\quad {\Sigma }^{\left(
1\right) }\left( \beta \right) =\boldsymbol{\psi \,}^{\left( 1\right)
}\left( \beta _{+}\right) +\boldsymbol{\psi \,}^{\left( 1\right) }\left(
\beta _{-}\right)   \tag{A5}
\end{equation}%
where $\boldsymbol{\psi \,}\left( z\right) =d\ln \Gamma \left( z\right) /dz$
is the digamma (psi) function and $\boldsymbol{\psi \,}^{\left( 1\right)
}\left( z\right) =d\boldsymbol{\psi \,}\left( z\right) /dz$ is the trigamma
function \cite{abr}, printed in a boldface type to differ from the Dirac
eigenspinor in Sec. 2, one can write%
\begin{equation}
\frac{\partial }{\partial \lambda }B\left( \beta _{+},\beta _{-}\right) =%
\frac{\beta _{1}}{\gamma }B\left( \beta _{+},\beta _{-}\right) {\Delta }%
\left( \beta \right)   \tag{A6a}
\end{equation}%
\begin{equation}
\frac{\partial ^{2}}{\partial \lambda ^{2}}B\left( \beta _{+},\beta
_{-}\right) =\left( \frac{\beta _{1}}{\gamma }\right) ^{2}B\left( \beta
_{+},\beta _{-}\right) \left[ {\Sigma }^{\left( 1\right) }\left( \beta
\right) +{\Delta }^{2}\left( \beta \right) \right]   \tag{A6b}
\end{equation}%
Finally, setting the parameter $\lambda =1$ and defining $\widetilde{\beta }%
_{\pm }=\beta _{\pm }\left( 1\right) $, one finds%
\begin{equation}
\int_{0}^{\infty }dx\,\frac{x\sinh 2\beta _{1}x}{\cosh ^{2\beta _{2}}\gamma x%
}=\frac{2^{2\beta _{2}}}{8\gamma ^{2}}B\left( \widetilde{\beta }_{+},%
\widetilde{\beta }_{-}\right) {\Delta }\left( \widetilde{\beta }\right)
\tag{A7a}  \label{A7a}
\end{equation}%
\begin{equation}
\int_{0}^{\infty }dx\,\frac{x^{2}\cosh 2\beta _{1}x}{\cosh ^{2\beta
_{2}}\gamma x}=\frac{2^{2\beta _{2}}}{16\gamma ^{3}}B\left( \widetilde{\beta
}_{+},\widetilde{\beta }_{-}\right) \left[ {\Sigma }^{\left( 1\right)
}\left( \widetilde{\beta }\right) +{\Delta }^{2}\left( \widetilde{\beta }%
\right) \right]   \tag{A7b}  \label{A7b}
\end{equation}%
for Re\thinspace $\widetilde{\beta }_{\pm }>0$.

Notice that because $\Gamma \left( z\right) $ has simple poles at $z=-n$
with residue $\left( -1\right) ^{n}/n!$ \cite{abr}, one has $\Gamma \left(
z\right) \simeq z^{-1}$ for $z\simeq 0$. As a consequence,%
\begin{equation}
B\left( z_{1},z_{2}\right) \simeq \frac{1}{z_{1}}+\frac{1}{z_{2}},\text{%
\quad for\quad }z_{1}\simeq z_{2}\simeq 0  \tag{A8a}  \label{A8a}
\end{equation}%
and%
\begin{equation}
\boldsymbol{\psi \,}\left( z\right) \simeq -\frac{1}{z},\text{\quad }%
\boldsymbol{\psi \,}^{\left( 1\right) }\left( z\right) \simeq \frac{1}{z^{2}}%
,\text{\quad for\quad }z\simeq 0  \tag{A8b}  \label{A8b}
\end{equation}%
On the other hand, because $\ln \Gamma \left( z\right) \simeq z\ln z$ for $%
z>>1$ \cite{abr}, one finds%
\begin{equation}
\boldsymbol{\psi \,}\left( z\right) \simeq \ln z,\quad \boldsymbol{\psi \,}%
^{\left( 1\right) }\left( z\right) \simeq \frac{1}{z},\text{\quad for\quad }%
z>>1  \tag{A9}  \label{A9}
\end{equation}

\bigskip

\bigskip

\bigskip

\bigskip

\bigskip

\bigskip

\noindent \textbf{Acknowledgments}

The authors gratefully acknowledge an anonymous referee for his/her valuable
comments and suggestions. This work was supported in part by means of funds
provided by CNPq.

\newpage

\begin{figure}[th]
\begin{center}
\includegraphics[width=14cm, angle=0]{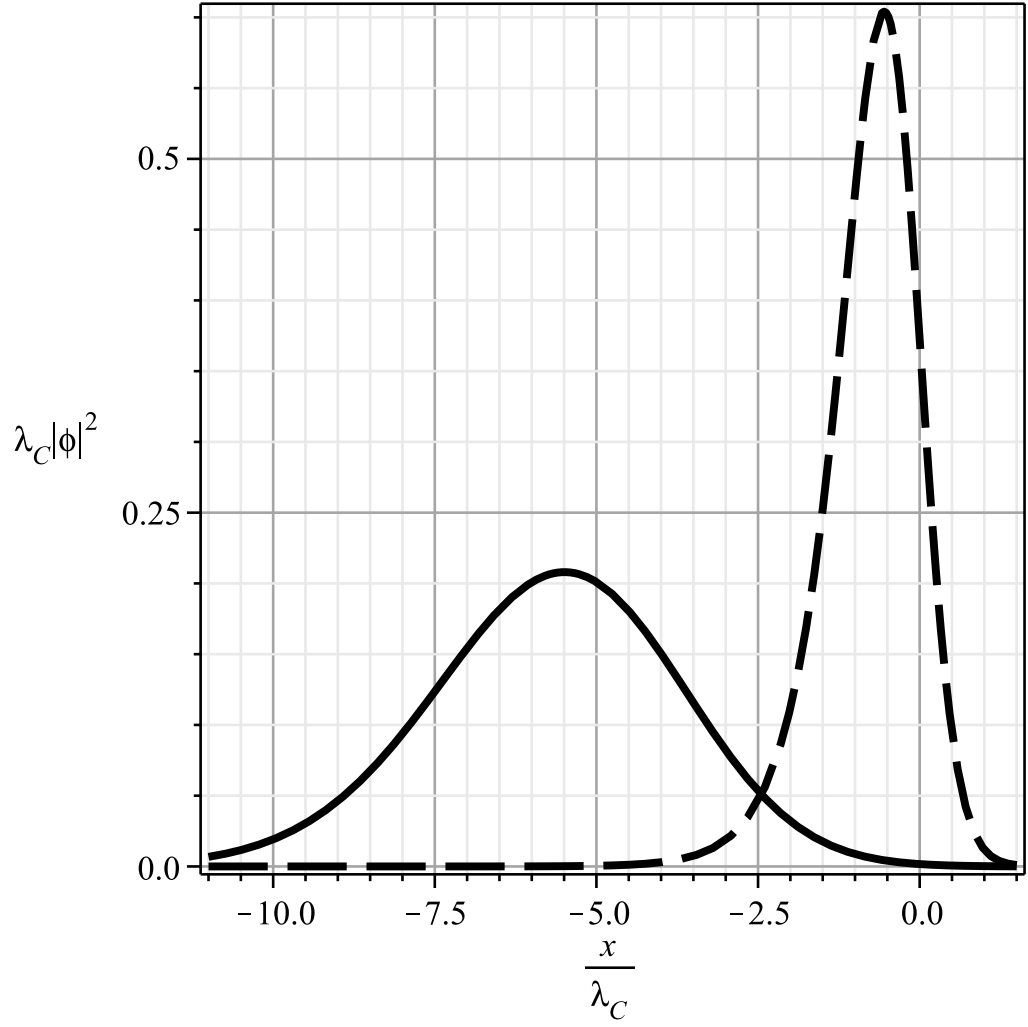}
\end{center}
\par
\vspace*{-0.1cm}
\caption{Position probability density for the isolated solution with $%
v_{0}/mc^{2}=2$ and $\protect\theta =3\protect\pi /8$. The continuous line
for $\hbar \protect\gamma /mc=1/10$, and the dashed line for $\hbar \protect%
\gamma /mc=1$. $\protect\lambda _{C}=\hbar /mc$ denotes the Compton
wavelength of the fermion.}
\label{Fig1}
\end{figure}

\begin{figure}[th]
\begin{center}
\includegraphics[width=14cm, angle=0]{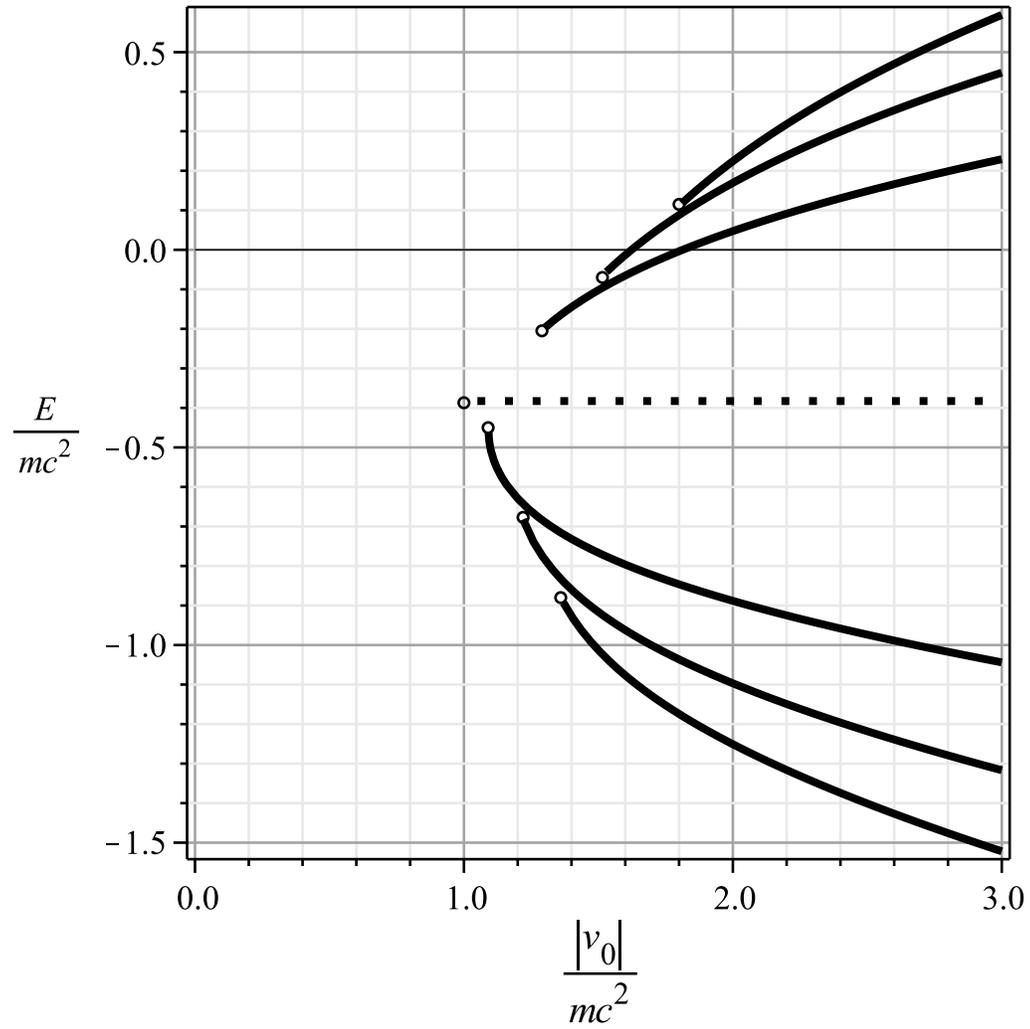}
\end{center}
\par
\vspace*{-0.1cm}
\caption{Energy levels for the three lowest quantum numbers ($n=1,2,3$ for $%
v_{0}>0$) with $\protect\theta =3\protect\pi /8$ and $\hbar \protect\gamma %
/mc=1/10$. The innermost curves are related to the lowest quantum numbers.
The dotted line is related to the isolated solution.}
\label{Fig2}
\end{figure}

\begin{figure}[th]
\begin{center}
\includegraphics[width=14cm, angle=0]{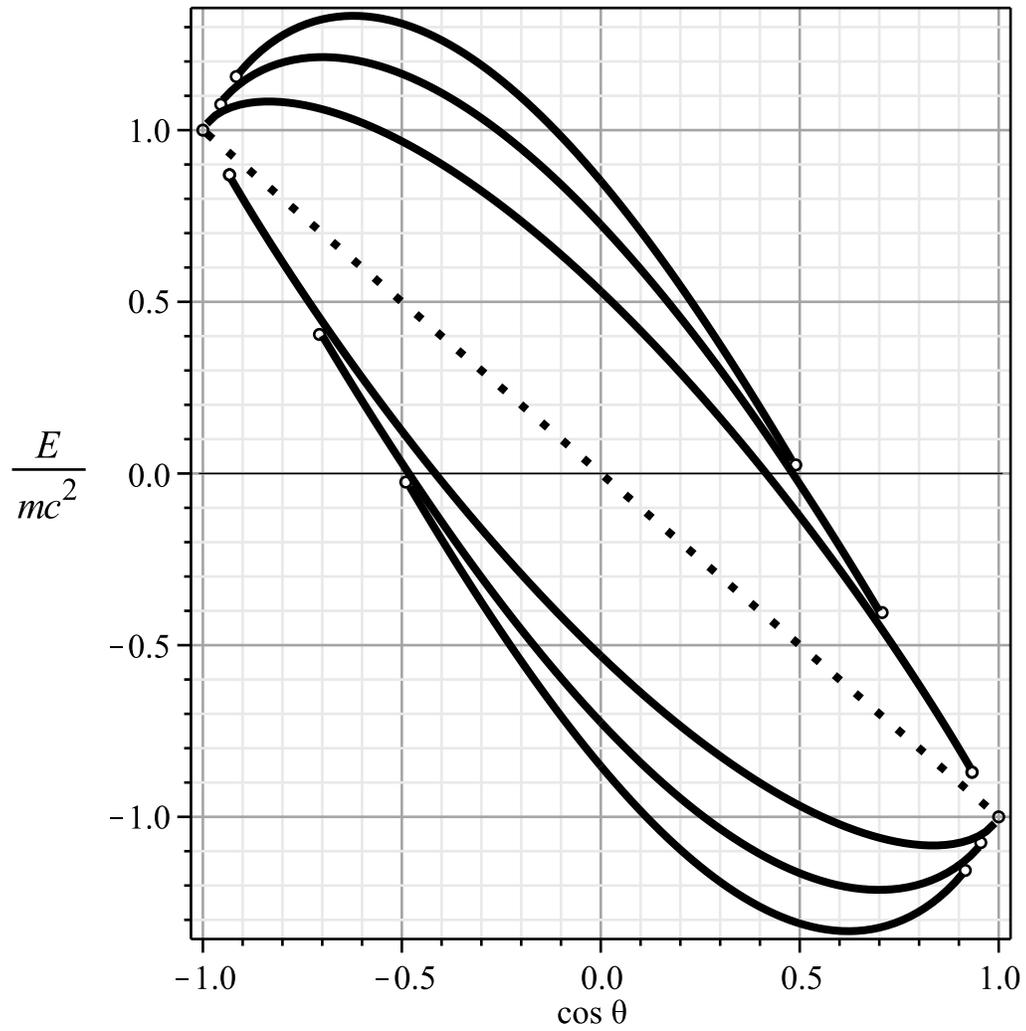}
\end{center}
\par
\vspace*{-0.1cm}
\caption{The same as Figure 2, for $v_{0}/mc^{2}=2$ and $\hbar \protect%
\gamma /mc=1/10$.}
\label{Fig3}
\end{figure}

\begin{figure}[th]
\begin{center}
\includegraphics[width=14cm, angle=0]{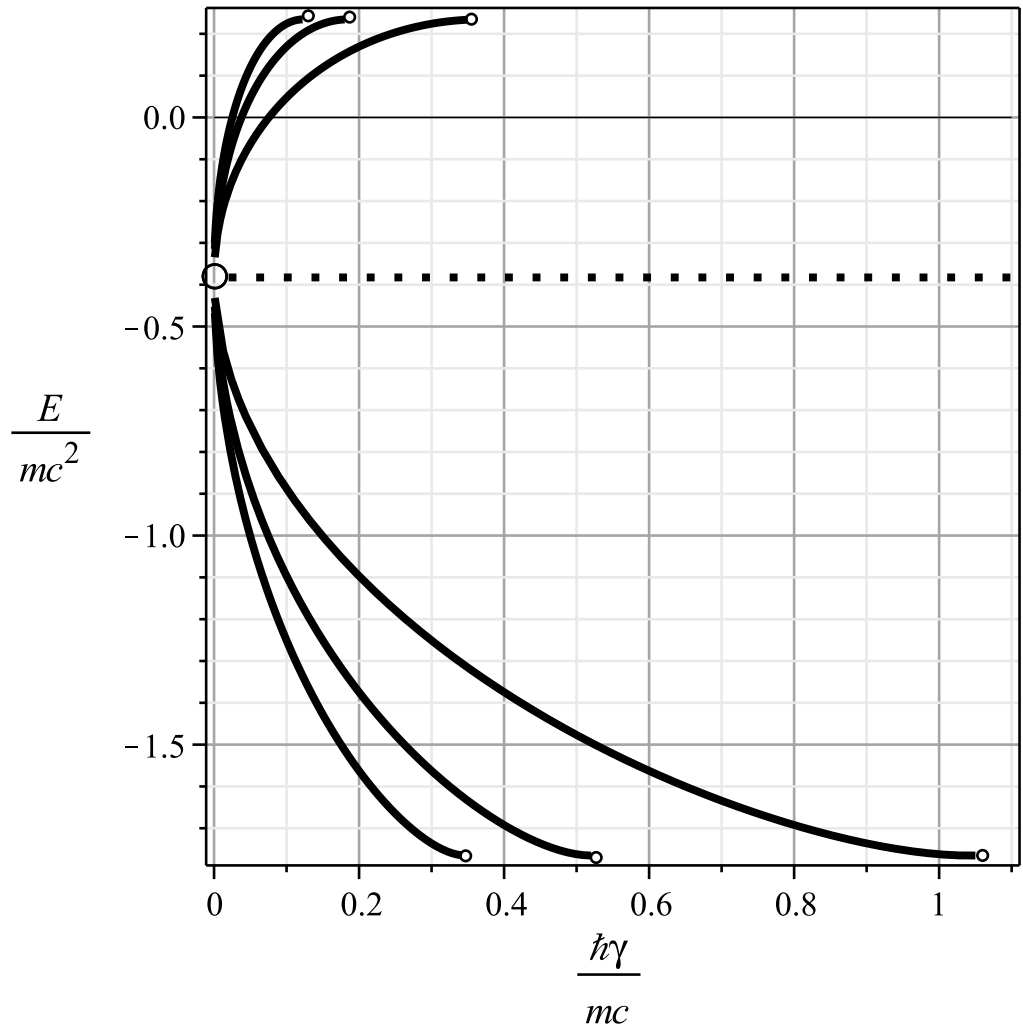}
\end{center}
\par
\vspace*{-0.1cm}
\caption{The same as Figure 2, for $\protect\theta =3\protect\pi /8$ and $%
v_{0}/mc^{2}=2$.}
\label{Fig4}
\end{figure}

\begin{figure}[th]
\begin{center}
\includegraphics[width=14cm, angle=0]{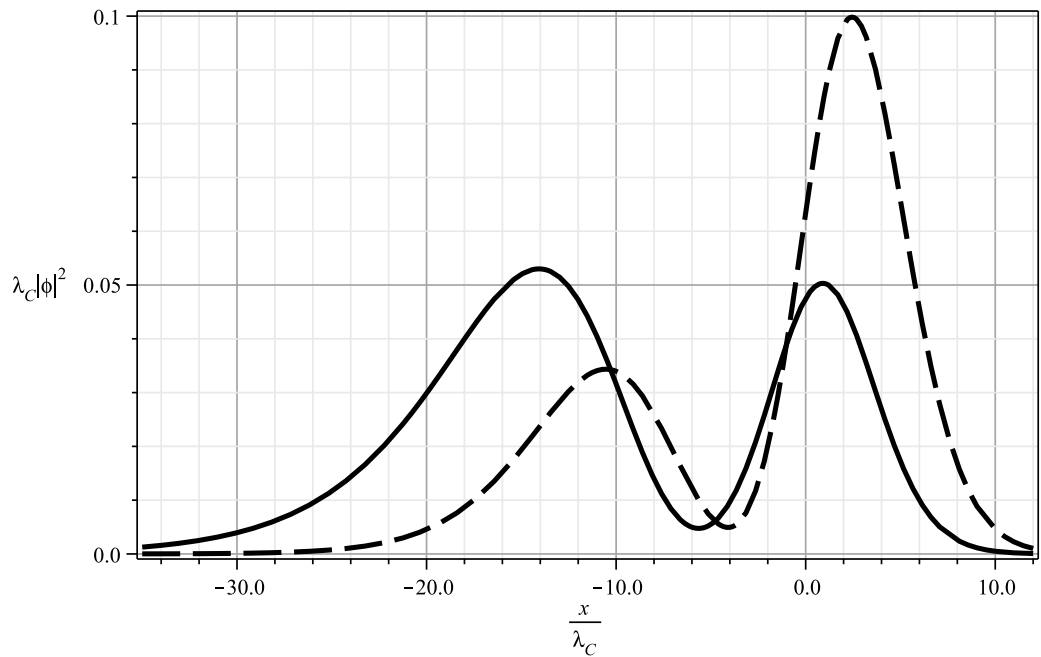}
\end{center}
\par
\vspace*{-0.1cm}
\caption{Position probability density for the Sturm-Liouville solution with $%
n=1$, $v_{0}/mc^{2}=2$, $\hbar \protect\gamma /mc=1/10$ and $\protect\theta %
=3\protect\pi /8$. The continuous line for $E/mc^{2}=+0.047$, and the dashed
line for $E/mc^{2}=-0.888$. $\protect\lambda _{C}=\hbar /mc$ denotes the
Compton wavelength of the fermion.}
\label{Fig5}
\end{figure}

\end{document}